# Tunable Extended Magnetic Non-Fermi Liquid in Graphene Moiré Heterostructures

Yongqin Xie[1,2†], Jian Wang[3†], Moyu Chen[1†], Chen Zhao[4], Fanqiang Chen[1], Qiao Li[1], Sicheng Chen[1], Jiao Xie[1], Kenji Watanabe[5], Takashi Taniguchi[6], Jin-hua Gao[4], Rui Wang[1], Shi-Jun Liang[1], Chunming Yin[3*], Bin Cheng[1,7*], Feng Miao[1,8*]

[1]Institute of Brain-Inspired Intelligence, National Laboratory of Solid State Microstructures, School of Physics, Collaborative Innovation Center of Advanced Microstructures, Nanjing University, Nanjing 210093, China.

[2]Hefei National Laboratory, Hefei 230088, China.

[3]CAS Key Laboratory of Microscale Magnetic Resonance and Department of Modern Physics, University of Science and Technology of China, Hefei 230026, China.

[4]School of Physics, Huazhong University of Science and Technology, Wuhan 430074, China.

[5]Research Center for Functional Materials, National Institute for Materials Science, 1-1 Namiki, Tsukuba 305-0044, Japan.

[6]International Center for Materials Nanoarchitectonics, National Institute for Materials Science, 1-1 Namiki, Tsukuba 305-0044, Japan.

[7]Institute of Interdisciplinary Physical Sciences, School of Science, Nanjing University of Science and Technology, Nanjing 210094, China.

[8]Jiangsu Physical Science Research Center, Nanjing 210093, China.

† These authors contribute equally to this work.

*Correspondence Email: Chunming@ustc.edu.cn; bincheng@njust.edu.cn; miao@nju.edu.cn

**Abstract:**

**Exploring exotic quantum metallic states beyond Landau's Fermi liquid theory remains a central focus in condensed matter physics. Such non-Fermi liquid (NFL) behavior is mostly observed near quantum criticality, yet growing attention is directed toward extended NFL phases with intrinsic quantum fluctuations rooted in the extended ground state. While these extended NFL states have been previously reported only in a limited set of *d*- and *f*-electron systems, realizing a single, highly tunable platform capable of exhibiting multiple resistance exponent $\alpha$ values is essential for uncovering the connection between the resistance exponent and the dominant quantum fluctuations coupled to quasiparticles. However, corresponding experimental progress remains elusive. Here, we report the observation of tunable extended non-Fermi liquid behavior in twisted double bilayer graphene encapsulated by aligned hBN layers. This NFL phase spans a broad range of carrier densities and exhibiting a carrier density–dependent resistance exponent. Combined with temperature-dependent resistance, magnetotransport and differential resistance measurements, these findings support a scenario where strong quantum fluctuations emerge from the interplay between localized and itinerant carriers. Our work establishes a highly tunable platform beyond conventional frameworks to investigate the organizing principles of non-Fermi liquid physics manifested in diverse behaviors.**

Landau's Fermi liquid theory based on the concept of coherent quasiparticles has been the fundamental framework for describing low-temperature electric transport properties of metals for more than half a century[1,2]. One of the distinguishing features of Fermi liquid is the $T^2$ scaling of the resistance at low temperatures, which is a direct result of the $T$-independent and short-range weak interactions between quasiparticles[3]. This framework based on Landau's quasiparticles, however, has been severely challenged by the emergence of non-Fermi liquid phases, in which the coherence of quasiparticles is destroyed[3-8]. Such breakdown of quasiparticles usually accompanies the emergence of quantum criticality in metals[3,6-10], where the strong quantum fluctuations of the order parameters give rise to soft collective boson modes that can couple to the elementary excitations near the Fermi surface (i.e., the quasiparticles), and thus lead to strong damping of the quasiparticles' lifetime[6]. Even more intriguing are the so-called extended NFL phases, which persist over wide regions of the phase diagram, far from any apparent quantum critical point. To date, the extended NFLs have been only reported in a few 3D d- and f-electron systems such as heavy Fermion systems[11], MnSi[12,13] and $ZrZn_2$[14]. These systems exhibit different resistance exponents arising from distinct underlying mechanisms, reflecting the diversity of low-energy quantum fluctuations involved. Realizing a single, tunable material system that can exhibit continuously variable resistance exponents would provide a concrete pathway toward establishing a unified understanding of these varied NFL behaviors.

In the past few years, two-dimensional (2D) twisted moiré heterostructures have emerged as highly tunable solid-state platforms for exploring exotic correlated states[15-29]. Recent theories suggest that these phenomena, including unconventional superconductivity, magnetism, and strange metallicity, may be understood within a unified framework based on the interplay between localized and itinerant carriers[30-34]. Indeed, recent experiments indicate that this interplay play an important role in the emergence of superconductivity and magnetism in moiré systems and multilayer graphene[35-37], yet how it may give rise to non-Fermi-liquid behavior remains largely unexplored. Here, we report the observation of an extended non-Fermi liquid state with a tunable resistance exponent in twisted double bilayer graphene encapsulated by aligned hBN layers. This NFL state emerges on the hole-doped side of a correlated insulating phase at the charge neutrality point (CNP), where electron localization dominates. Both the CNP insulating phase and the NFL regime exhibit clear magnetic hysteresis, while the latter additionally shows an unconventional resistance peak in the $R$–$T$ curves. Combined with magnetotransport and differential resistance measurements, these results reveal an enhanced interplay between localized and itinerant carriers, suggesting that their coexistence in the NFL regime generates strong quantum fluctuations that destroy quasiparticle coherence.

**Transport signatures of strong electronic correlations**

Our TDBG sample was fabricated by stacking two pieces of Bernal bilayer graphene in the AB-BA manner with a twisted angle $\theta$ = 1.26°, forming moiré patterns consisting of three distinct regions of stacking order ABBA, ABCB and ABAC (see Figs. 1(a) and 1(b)). Two hexagonal boron nitride (hBN) layers were inserted between the TDBG sample and dual graphite gates, so that we can independently control the perpendicular displacement field $D$ and the carrier density $n$ following the equations $D = (C_{tg}V_{tg} - C_{bg}V_{bg})/2$ and $n = (C_{tg}V_{tg} + C_{bg}V_{bg})$, in which $C_{tg}$ ($C_{bg}$) represents the capacitance between TDBG and top (bottom) gate. Notably, we intentionally aligned both the top and bottom hBN layers to the adjacent graphene sheets in the process of transfer. Theoretical calculations using the single-particle continuum model approximation show that the TDBG with such double-aligned configuration can achieve isolated and narrower flat bands (a bandwidth of ~13 meV for the first valance band) at zero electric field compared to the cases with single-aligned hBN or non-aligned hBN that have been widely studied (see Fig. 1(c) and Supplemental Material Fig. 1). In particular, the flattened region near the upper edge of the first valence band in the double-aligned configuration can enhance electronic correlations and lead to a large effective carrier mass at low carrier densities.

We then performed the transport measurements to explore the strongly correlated behaviors in this previously unidentified double-aligned graphene moiré heterostructure. As shown in Fig. 1(d), the longitudinal resistances ($R_{xx}$) as a function of $n$ and $D$ at zero magnetic field and base temperature 1.6 K present strong resistance peaks at the first moiré gaps at $\pm n_s = 3.7 \times 10^{12}\ cm^{-2}$, corresponding to four electrons (or holes) per moiré unit cell. In contrast to the results in previously reported TDBG devices[21-23], the integer correlated insulators on the electron side are absent, which can be attributed to the significantly changed band structures of the device with a double-aligned configuration (see Supplemental Material Fig. 2). The asymmetric features of $R_{xx}$ between the positive and negative electric displacement fields can be attributed to the structural asymmetry of the device, stemming from the inevitable inequality between top and bottom graphene-hBN alignments. Additionally, we note a resistance peak at small displacement field ($|D| < 0.1$ V/nm) at CNP, which is latter identified as a correlated insulating state—distinct from the semimetallic behavior observed in previously reported TDBG devices[21-23,38,39]. In proximity to this resistance peak on the hole-doped side, we observe an unconventional and sizeable region exhibiting higher resistance than that of a conventional metal (e.g., $|n| > n_s$). These phenomena can be attributed to the strong correlation in the double-aligned configuration, consistent with the flat valence band illustrated in the calculated band structures shown in Fig. 1(c). Notably, we observe similar electric transport behaviors in several devices with $\theta = 0.97° \sim 1.26°$ and here we focus on the device with $\theta = 1.26°$ (see Supplemental Material Fig. 3-6 for data of other devices). For comparison, we measured an ABBA-stacked TDBG device (D4) with $\theta = 1.29°$ without graphene-hBN alignment, which exhibits transport behaviors notably different from devices with double-aligned configuration (see Supplemental Material Fig. 7), but similar to previously reported TDBG devices[21-23].

To better understand the unconventional phenomena near CNP, we kept the displacement field at zero and applied a perpendicular magnetic field $B_\perp$, obtaining the longitudinal and Hall resistance as a function of $n$ and $B_\perp$ at $D = 0$ V/nm as shown in Figs. 1(e) and 1(f). Clear Landau levels originate from moiré miniband edge with charge densities $|n| > n_s$ at both electron and hole side at a small magnetic field ( $B_\perp \sim 0.5$ T ), demonstrating the high quality of our TDBG device (also see Supplemental Material Fig. 8 for the AFM image of the heterostructure). While filling the first conduction band, a quantum Hall state with Chern number $C = 6$ originate from CNP at small $B_\perp \sim 0.5$ T and persist to a large magnetic field (see Figs. 1(e), 1(f) and Supplemental Material Fig. 9). This $C = 6$ state can exist in a range of $D$, and will transit to topologically trivial band insulating states when $|D| > 0.25\ \mathrm{V/nm}$. We consider this $C = 6$ state may be a valley Chern insulator[40], which emerges with the assistance of a small $B_\perp$ and can be stable in a large $B_\perp$.

While filling the first flat valence band, however, quantum oscillations are absent even with $B_\perp$ up to 12T (see Figs. 1(e), 1(f) and Supplemental Material Fig. 10). In general, quantum oscillations occur in metals as the cyclotron energy $E_c$ is larger than the thermal energy $k_B T$ and the Landau quantization of itinerant fermions is facilitated.

Another prerequisite for the emergence of quantum oscillations is the existence of well-defined quasiparticle wavefunction, which remains coherent throughout an entire orbit of the Fermi surface under the magnetic field. In this sense, the absence of Landau levels suggests the disappearance of cyclotron orbits' coherence, which could be attributed to an extremely large Fermi surface and the strong enhancement of quasiparticle effective mass, or even the breakdown of coherence of quasiparticles.

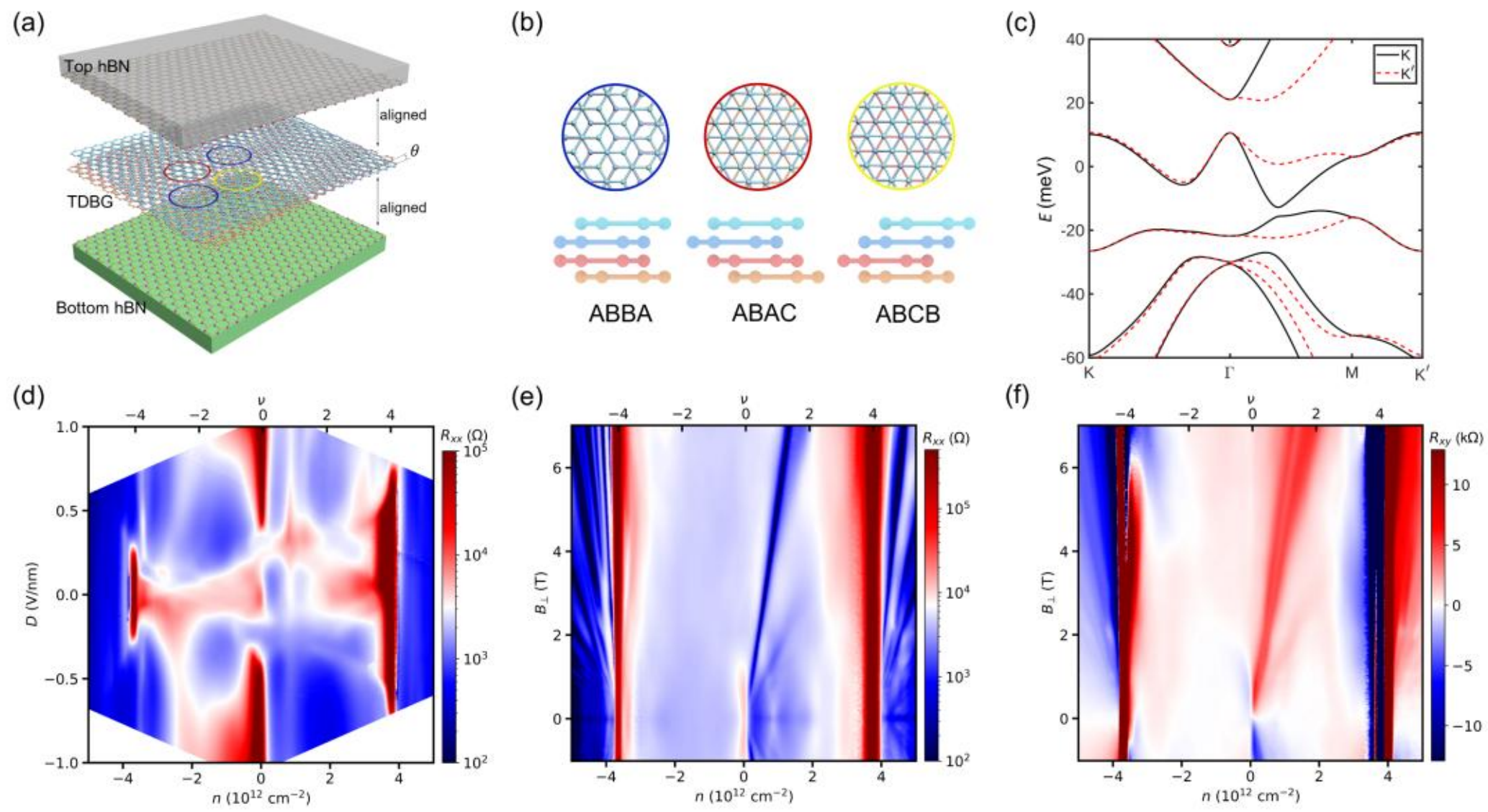

**Figure 1. Transport in a 1.26° double-aligned TDBG device.** (a) Schematic of an ABBA-stacked TDBG device with adjacent hBN and graphene sheets aligned. The circles of different colors represent the regions of different stacking orders. (b) Top and side views of the different stacking orders marked in (a). (c) Calculated band structure for $\theta = 1.26°$ ABBA-stacked TDBG with double-aligned hBN layers at zero electric displacement field. (d) Longitudinal resistance as a function of carrier density $n$ and displacement field $D$ at $T$ = 1.6 K. (e),(f) $R_{xx}$ and $R_{xy}$ as a function of carrier density $n$ and perpendicular magnetic field $B_\perp$ at $D = 0\mathrm{V/nm}$. Landau levels are not observed when filling the first valence band.

### *T*-dependent resistivity and magnetic hysteresis

We then investigate the temperature dependence of longitudinal resistance $R_{xx}$ at zero displacement field and magnetic field. Figure 2(a) shows $R_{xx}$ as a function of carrier density $n$ and temperature $T$ at $D$ = 0. The temperature dependence of resistance at very small carrier densities exhibits insulating behaviors and follows $R_{xx} \propto \exp\left(\frac{T_{ES}}{T}\right)^{\alpha}$, with $\alpha = 1/2$ and $T_{ES}$ being the characteristic temperature (the inset of Fig. 2(b)). This value of $\alpha$ is consistent with the Efros-Shklovskii (ES) variable range hopping mechanism[36,39,41,42]. Such localized state will transition into delocalized metallic states at around 8 K. To characterize the localization behaviors of the carriers, we extract the localization length $\xi$ by fitting the equation $T_{ES} = \frac{(2.8e^2)}{4\pi\varepsilon_0\varepsilon k_B\,\xi}$, where $\varepsilon_0$ and $\varepsilon$ denote the vacuum electric permittivity and the dielectric constant of hBN, respectively. Figure 2(b) shows that the $\xi$ near CNP is comparable to the length of channel (~1 μm), which is about 100 times the wavelength of moiré superlattice.

Besides, we observe a clear hysteresis in resistance at the CNP and zero displacement field ($D$ = 0 V/nm) when sweeping the out-of-plane magnetic field $B_{\perp}$ forward and backward (see Fig. 2(c)). This hysteresis gradually weakens with increasing temperature and becomes nearly indiscernible around 5 K (see Supplemental Material Fig. 11), close to the delocalization transition temperature of localized states, implying that carrier localization plays a key role in the observed magnetic hysteresis.

We then dope holes into the insulating state through electric gating, and Fig. 2(e) shows the temperature dependence of $R_{xx}$ at varied hole densities. Above 20 K, $R_{xx}$ is linear in temperature with $\partial R_{xx}/\partial T \approx 100\ \Omega/\mathrm{T}$, which is similar to the previously reported strange metal phases in twisted bilayer graphene and twisted double bilayer graphene[15,21,43,44]. As the temperature decreases, the resistance reaches a minimum at a characteristic temperature $T_K'$ and follows a $-\ln T$ scaling (see Supplemental Material Fig. 12). When the temperature is further lowered, $R_{xx}$ deviates this scaling and decreases dramatically below a characteristic temperature $T^*$, at which the $R$-$T$ curves exhibiting a peak or a kink. To illustrate the temperature-dependent behaviors more intuitively, we extract the $n$-$T$ mapping of the first derivative of resistance ($dR_{xx}/dT$) as shown in Fig. 2(d). The negative $dR_{xx}/dT$ signals a clear deviation toward the hole-doped side, and $T^*$ marks the boundary between regions with metallic and insulating behavior. This rare negative d$R$/d$T$ in metal, together with the observed $-\ln T$ scaling, is commonly associated with scattering of itinerant carriers by local moments[37,45-47].

Moreover, applying in-plane magnetic fields can tune these temperature-dependent behaviors (see Supplemental Material Fig. 13), implying the mechanism related to the magnetism.

Besides the signatures in thermal activation behavior, we also observe unconventional magnetic hysteresis behavior in hole doping region when sweeping $B_{\perp}$ (see Fig. 2(f)). By measuring at different positions in the *n*-*D* mapping, we found that pronounced hysteresis emerges within the unconventional metallic region adjacent to the CNP, whereas it is absent outside this regime (see Supplemental Material Fig. 14). This magnetic hysteresis in the metallic state, similar to that observed at the CNP, may originate from localized carriers.

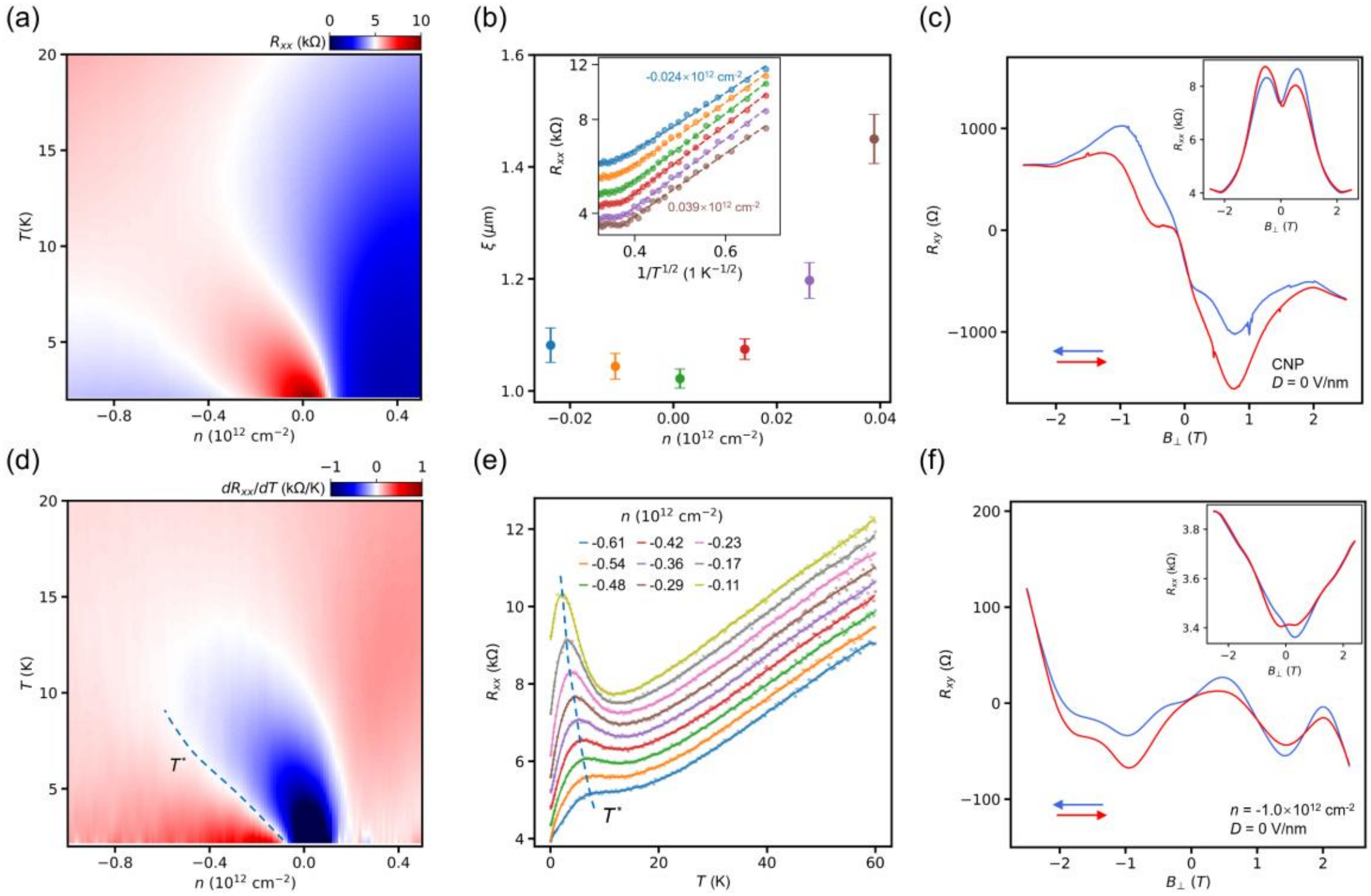

**Figure 2. Signatures of ES-VRH conduction, nagetive d*R*/d*T* and magnetism.** (a) Longitudinal resistance $R_{xx}$ as a function of carrier density $n$ and temperature $T$ at $D$ = 0 V/nm. (b) Doping dependence of the localization length $\xi$ extracted from ES variable range hopping model. $R_{xx}$ (log scale) versus $T^{-1/2}$ with varied carrier densities near CNP at $D$ = 0 V/nm (the inset). (c) Hysteresis in $R_{xy}$ and $R_{xx}$ (the inset) as a function of perpendicular magnetic field at CNP at 2 K. The forward and backward sweeping of magnetic field are shown in red and blue, respectively. (d) $dR_{xx}/dT$ as a function of $n$

and $T$ at $D$ = 0 V/nm. The blue dashed line denotes the characteristic temperature $T^*$. (e) $T$-dependence of $R_{xx}$ at different holes doping at $D$ = 0 V/nm. $T^*$ is marked by a dashed line as a guide to the eye. (f) Hysteresis of $R_{xy}$ and $R_{xx}$ (the inset) at $n$ = -1.0×$10^{12}$ $cm^{-2}$ and $D$ = 0 V/nm at 2 K. The forward and backward scans are shown in red and blue, respectively.

**Extended non-fermi liquid and reentrant insulating states at low temperature**

To study the low-temperature behaviors, we measured the resistance down to 67 mK while varying $n$ and $D$. By fitting the formula $R = AT^{\alpha} + R_0$, the exponent $\alpha$ is obtained at different fillings. Remarkably, we observe a substantial region of non-Fermi liquid having $R$-$T$ relation deviated from $T^2$ scaling law, as highlighted in Fig. 3(a). Outside this region, the system evolves into either an insulating state or a Fermi-liquid regime at low temperatures. Representative $R$–$T$ curves at several different fillings marked in Fig. 3(a) are shown in Fig. 3(b). While the electron-doped side follows Fermi-liquid behavior (red squares), the extracted $\alpha$ values deviate from 2 in the hole-doped region. Particularly, the exponent $\alpha$ at the filling $n$ and displacement field $D$ represented by blue point in Fig. 3(a) is fitted to be 1, with a slope of 288 Ω/K, demonstrating the emergence of low-temperature strange metal phase[15]. We then performed $T$-dependence measurements at $D$ = 0 with high-resolution carrier doping. The extracted $n$-$T$ phase diagram exhibits an extended NFL region occupying a substantial range of $n$, with $\alpha$ values varying with different $n$ (Figs. 3(c) and 3(d); see also Supplemental Material Fig. 15 for details). This non-Fermi liquid manifests as a stable low-temperature phase extending over a wide parameter range without signatures of quantum criticality. We consider that the interplay between localized and itinerant carriers, existing over extended range may give rise to strong quantum fluctuations that destroy the coherence of quasiparticle, consistent with the observed absence of quantum oscillations.

Moreover, we observe reentrant insulating (RI) states embedded within the extended non-Fermi liquid regime (Fig. 3b-d). Typically, increasing the density of itinerant carriers enhances metallic behavior. In our case, continuously doping holes drives the system from a non-Fermi liquid state back into an insulating phase, highlighting the complex interplay between localized and itinerant electrons in the system.

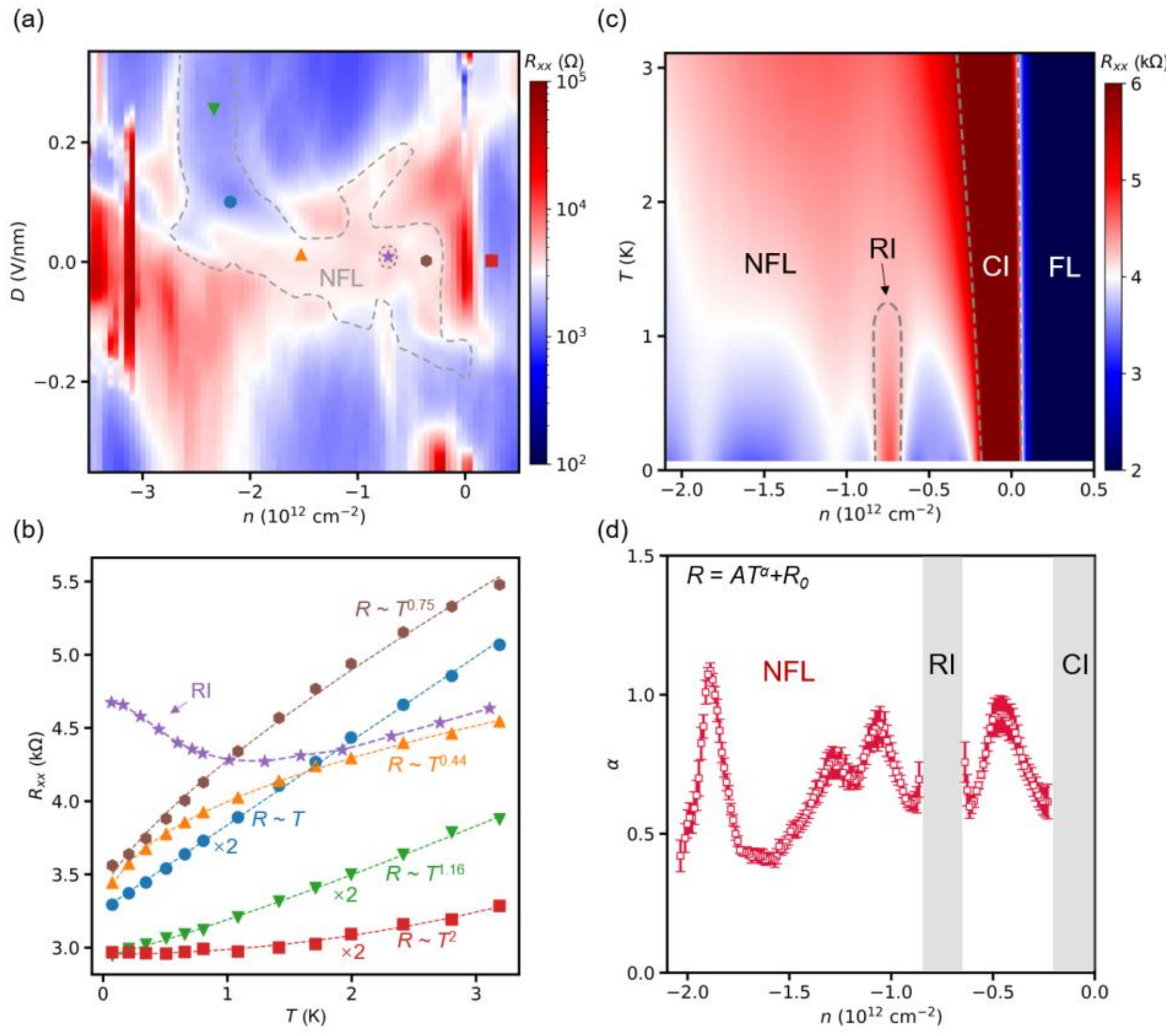

**Figure 3. Extended non-Fermi liquid.** (a) Longitudinal resistance $R_{xx}$ as a function of carrier density $n$ and displacement field $D$ at $T = 67$ mK. (b) $R$-$T$ behaviors at different positions marked by different markers in (a). (c) $R_{xx}$ as a function of $n$ and $T$ at $D = 0$ V/nm. (d) Resistance exponent $\alpha$ of NFL as a function of carrier density $n$. NFL, FL, CI, and RI denote non-Fermi liquid, Fermi liquid, correlated insulator, and reentrant insulator, respectively.

## Bias-current-induced phase transition

To gain further insights into the coexistence and interplay of localized and itinerant carriers in hole-doped regime, we perform differential resistance measurements (d$V$/d$I$ versus $I_{dc}$) across a range of filling factors and temperatures as presented in Fig. 4. At the CNP, the differential resistance of the insulating state drops sharply with increasing bias current at low temperatures (see Fig. 4(a)), indicating a current-induced delocalization of localized charge carriers. Upon hole doping, the differential resistance also changes rapidly with increasing bias current at low temperatures, both at -0.11 and $-1.75\times10^{12}$ cm$^{-2}$ (see Fig. 4(b) and 4(c)). To further characterize the current-induced evolution, we extract the d$V$/d$I$-$T$ curves at varying bias currents (Figs. 4(d)-4(f)). The insulating behavior at the CNP is progressively suppressed with increasing bias current (Fig. 4(d)). At $-0.11\times10^{12}$ cm$^{-2}$, the unconventional metallic state emerging at small currents and low temperatures is destroyed by large currents, reverting to an insulating behavior (Fig. 4(e)). At $-1.75\times10^{12}$ cm$^{-2}$, the non-Fermi liquid behavior gradually evolves toward a Fermi-liquid-like behavior as the bias current increases (Fig. 4(f)). The bias-current dependence of the metallic states reveals a close connection to current-induced delocalization, implying the coexistence and interplay between localized and itinerant carriers. As the bias current increases, the localized carriers become progressively delocalized, thereby weakening their coupling to itinerant carriers and driving the unconventional states toward behavior consistent with the Fermi-liquid theory.

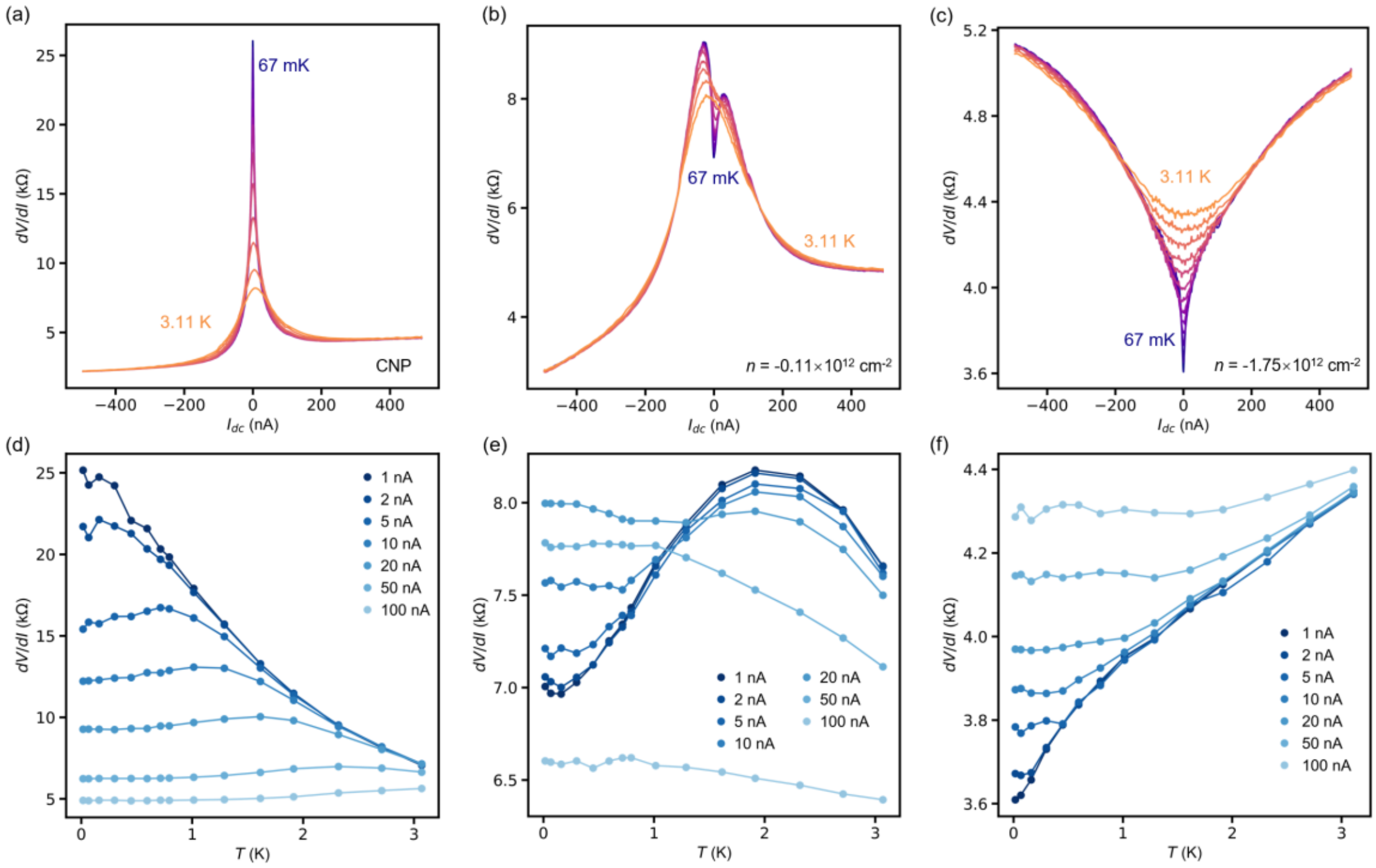

**Figure 4. Differential resistance measurements.** (a)-(c) The longitudinal differential resistance $dV/dI$ vs bias current $I_{dc}$ at various temperature $T$ for the CNP, $-0.11\times10^{12}$ cm$^{-2}$ and $-1.75\times10^{12}$ cm$^{-2}$. (d)-(f) Corresponding temperature-dependent differential resistance at selected bias currents.

**Discussion and conclusion**

In the preceding sections, our experimental data indicate a tunable, extended NFL behavior in the resistance, described by $R = AT^{\alpha} + R_0$, with the exponent $\alpha$ varying between ~0.4 and ~1.2 as a function of $n$. On the hole-doped side near the CNP, the system displays a peculiar temperature dependence at finite temperatures, characterized by a peak or kink in the $R$–$T$ curves. Notably, similar temperature dependence has been observed previously in rhombohedral multilayer graphene and was attributed to the scattering of itinerant electrons by local moments formed from spatially confined isospin excitations[37]. In our experiments, the results of temperature-dependent resistance, magnetic hysteresis, and differential resistance measurements also imply the interplay between magnetically ordered localized carriers and itinerant carriers. The coexistence of magnetic moments and re-entrant insulating behavior further suggests the emergence of quantum fluctuations in both magnetic and charge order parameters, which couple to quasiparticles and destroy their coherence, giving rise to an extended non-Fermi liquid phase and the absence of quantum oscillations.

Going a step further, our observations may be discussed within a Kondo lattice–type framework, involving interactions between periodically arranged localized moments and itinerant carriers. The $-\ln T$ scaling in the resistance versus temperature (R–T) curve is a hallmark of Kondo scattering[45-47]. When the temperature drops below the characteristic temperature $T^*$, the resistance rapidly decreases with decreasing temperature, consistent with the formation of a coherent Kondo state[46-49]. The measurements under in-plane magnetic fields are consistent with the suppression of Kondo antiferromagnetic hybridization and the consequent disruption of the Kondo coherent state. Similarly, the bias current delocalizes local moments, thereby destroying the Kondo coherence. Notably, the residual magnetic hysteresis observed in the hole-doped regime may imply that the Kondo hybridization is of the undercompensated type, offering a unified framework for interpreting the low-temperature NFL behavior and the reentrant insulating state. At low carrier densities, localized orbitals can emerge due to the flattened valence band near the band edge. These localized states can be effectively described by a periodic Anderson model with emergent SU(4) isospin symmetry[30,31], owing to the negligible spin–orbit coupling in graphene. Upon hole doping, itinerant carriers couple to the local SU(4) moments via Kondo exchange, giving rise to an underscreened Kondo coherent state. In this regime, residual local moments may interact through RKKY or direct exchange mechanisms, leading to long-range antiferromagnetic (AFM) order (with parameter $m$) that coexists with Kondo screening (with parameter $\chi$)[50]. This coexistence accounts for the observed magnetic hysteresis in the Kondo hybridization regime. With further doping, increased carrier density suppresses both types of ordering and enhances their competition, giving rise to strong low-energy quantum fluctuations. These carriers can also be scattered by underscreened Kondo clouds, characterized by a spatial fluctuation of the Kondo order parameter $\delta\chi(r)$, which can lead to a finite relaxation time of carriers and drive a linear-in-$T$ resistivity[51]. This resembles a Yukawa-SYK model where electrons couple to random critical bosonic modes[52,53]. Moreover, extrinsic disorder enhances spatial fluctuations that break lattice translation symmetry, preserving their coupling to carriers relevant at low energies and thereby stabilizing NFL behavior[54]. Additionally, AFM spin wave fluctuations can also couple to carriers and may modify the exponent in the resistivity, explaining the observed $R \sim AT^{\alpha} + R_0$ with doping-dependent $\alpha$. Finally, at certain doping levels, a second-stage Kondo screening may occur between itinerant holes and residual moments[55,56], driving the system from an NFL to an insulating state—consistent with the observed reentrant insulating behavior.

In conclusion, we observed an extended NFL regime in ABBA-stacked twisted double bilayer graphene (TDBG) with double-aligned hBN layers, which can be effectively tuned via carrier density, vertical electric field, and external magnetic field. The observed NFL behavior, characterized by a continuously tunable exponent $\alpha$, can be attributed to interplay between itinerant and localized carriers within the flat bands. This interplay facilitates the coexistence of NFL behavior and long-range magnetic order, further giving rise to a reentrant insulating state and suppression of quantum oscillations, underscoring the role of quantum fluctuations in both magnetic and charge order parameters. Our findings establish graphene moiré heterostructures as a highly tunable platform for exploring extended NFL physics beyond the Landau framework, offering new insights into the interplay between localized and itinerant electrons in two-dimensional materials.

## Methods

**Device fabrication.** The encapsulated ABBA-stacked TDBG with alignment between graphene sheets and hBN layers were fabricated by a modified polymer-based cut-and-pick technique. Flakes of bilayer graphene, graphite and hBN (10-25 nm thick) were first mechanically exfoliated onto $SiO_2$/Si wafers, and chosen with optical microscopy and atomic force microscopy (AFM). Flakes of bilayer graphene and hBN are selected with long and straight edges, which correspond to the crystallographic axe (armchair or zigzag) with high likelihood. We then cut the bilayer graphene flake into two halves by using an AFM tip in contact mode for mechanic transfer. The transfer for fabricating van der Waals (vdW) heterostructures was done by using stamps consisting of poly (bisphenol A carbonate) (PC) film and polydimethylsiloxane (PDMS) to pick up top graphite and hBN at around 80℃, and then two halves of bilayer graphene, with a rotation of 180°+1.3°, at room temperature. Finally, we picked up the bottom hBN flake at 80℃ and transferred the vdW heterostructure onto the bottom graphite on $SiO_2$/Si substrate at around 135℃. In the process of transfer, the straight edges of top (bottom) bilayer graphene and top (bottom) hBN were aligned, which was confirmed by using an optical microscopy. The assembled vdW heterostructures were then shaped into a Hall bar geometry by using standard electron beam lithography (EBL) and dry etching in an inductively coupled plasma (ICP) system. The contact electrodes were deposited by a standard electron beam evaporation system of Cr/Pd/Au (5 nm/15 nm/30 nm). The temperature of the stacking process was kept below 150℃ during the fabrication.

**Electrical measurements.** We have two data sets obtained in an Oxford cryostat with a base temperature of ~ 1.6 K and a Bluefors dilution refrigerator down to ~ 60 mK, respectively. The transport measurements were performed using a lock-in amplifier (Stanford Research SR830) with an a.c. excitation current of 1 nA at 17.7 Hz, unless otherwise stated. Gate voltages are applied via Keithley 2400 Source Meters.

**Continuum Hamiltonian of ABBA-stacked Twisted Double Bilayer Graphene.**

We consider the ABBA-stacked twisted double bilayer graphene (TDBG) at twist angle $\theta = 1.26°$. The TDBG is composed by stacking an AB-stacked bilayer graphene (BLG) on top of another AB-stacked BLG with $B_1, A_2, A_3, B_4$ vertically aligned. Then we rotate the first and the second BLGs by $-\theta/2$ and $\theta/2$, respectively. We adopt the continuum model to describe the electronic band structure of AB-BA-stacked TDBG. The Hamiltonian of ABBA-stacked TDBG at small twist angle is

$$H = \begin{pmatrix} H_0(\mathbf{k_1}) & g^\dagger(\mathbf{k_1}) & & \\ g(\mathbf{k_1}) & H_0'(\mathbf{k_1}) & U^\dagger & \\ & U & H_0'(\mathbf{k_2}) & g(\mathbf{k_2}) \\ & & g^\dagger(\mathbf{k_2}) & H_0(\mathbf{k_2}) \end{pmatrix}$$

$$H_0(\mathbf{k}) = \begin{pmatrix} 0 & -\hbar v k_- \\ -\hbar v k_+ & \Delta' \end{pmatrix}$$

$$H_0'(\mathbf{k}) = \begin{pmatrix} \Delta' & -\hbar v k_- \\ -\hbar v k_+ & 0 \end{pmatrix}$$

$$g(\mathbf{k}) = \begin{pmatrix} \hbar v_4 k_+ & \gamma_1 \\ \hbar v_3 k_- & \hbar v_4 k_+ \end{pmatrix}$$

where $\mathbf{k}_l = R\left(\pm\frac{\theta}{2}\right)\left(\mathbf{k} - \mathbf{K}_\xi^{(l)}\right)$ with $\pm$ for $l = 1$ and 2, respectively and $k_\pm = \xi k_x \pm i k_y$. $\Delta' = 0.05\text{eV}$ and the band velocity of graphene monolayer $v$ is taken as $\hbar v/a = 2.1354\text{eV}$. $g$ is the interlayer coupling of the AB-stacked BLG and the parameter is set by $v_i = (\sqrt{3}/2)\gamma_i a/\hbar$ with $\gamma_1 = 0.4\text{eV}$, $\gamma_3 = 0.32\text{eV}$, $\gamma_4 = 0.044\text{eV}$. $U$ describes the moiré interlayer coupling between the twisted layers and is given by

$$U = \begin{pmatrix} u & u' \\ u' & u \end{pmatrix} + \begin{pmatrix} u & u'\omega^{-\xi} \\ u'\omega^{\xi} & u \end{pmatrix} e^{i\xi \mathbf{G}_1^{\mathrm{M}}\cdot\mathbf{r}} + \begin{pmatrix} u & u'\omega^{\xi} \\ u'\omega^{-\xi} & u \end{pmatrix} e^{i\xi(\mathbf{G}_1^{\mathrm{M}}+\mathbf{G}_2^{\mathrm{M}})\cdot\mathbf{r}}$$

where $\omega = e^{2\pi i/3}$, $u = 0.0797\text{eV}$, $u' = 0.0975\text{eV}$.

When the system is aligned on a hexagonal boron nitride (hBN) substrate, the interaction between the hBN substrate and the adjacent graphene layer breaks the sublattice symmetry and opens a band gap. We introduce the effect of the hBN substrate by setting the on-site energy of the sublattices in the adjacent graphene layer to be different to induce a band gap. The magnitude of the band gap is set by $\Delta = 0.053$eV.

We calculate the electronic band structure of the ABBA-stacked TDBG without hBN substrate, aligned with hBN substrate at bottom layer, and aligned with hBN substrates at both top and bottom layers using the continuum model introduced above.

**Acknowledgments**

This work was supported in part by the National Key R&D Program of China under Grant 2025YFA1411003, 2023YFF0718400 and 2023YFF1203600, the National Natural Science Foundation of China (12322407, 62034004), the Natural Science Foundation of Jiangsu Province (No. BK20233001), the Leading-edge Technology Program of Jiangsu Natural Science Foundation (BK20232004), the Strategic Priority Research Program of the Chinese Academy of Sciences (XDB44000000), Innovation Program for Quantum Science and Technology (2024ZD0300101), Fundamental and Interdisciplinary Disciplines Breakthrough Plan of the Ministry of Education of China (JYB2025XDXM120). F.M. and S.-J.L. would like to acknowledge support from the Fundamental Research Funds for the Central Universities (020414380227, 020414380240, 020414380242) and the e-Science Center of Collaborative Innovation Center of Advanced Microstructures. J. Y. and K. X. would like to acknowledge support from Zhejiang Provincial Natural Science Foundation of China under Grant No. LQN26A040023. K.W. and T.T. acknowledge support from the JSPS KAKENHI (Grant Nos. 21H05233 and 23H02052) and the World Premier International Research Center Initiative, MEXT, Japan. We thank Jianpeng Liu and Zhongkai Liu for fruitful discussions.

## Data availability:

The data that support the plots within this paper and other findings of this study are available from the corresponding authors upon reasonable request.

## Reference:

1 Landau, L. D. The theory of a Fermi liquid. *Soviet Physics Jetp-Ussr* **3**, 920-925 (1957).

2 Venema, L. *et al.* The quasiparticle zoo. *Nature Physics* **12**, 1085-1089 (2016).
3 Stewart, G. R. Non-Fermi-liquid behavior in d- and f-electron metals. *Reviews of Modern Physics* **73**, 797-855 (2001).
4 Lee, S.-S. Recent Developments in Non-Fermi Liquid Theory. *Annual Review of Condensed Matter Physics* **9**, 227-244 (2018).
5 Imada, M., Fujimori, A. & Tokura, Y. Metal-insulator transitions. *Reviews of Modern Physics* **70**, 1039-1263 (1998).
6 Löhneysen, H. v., Rosch, A., Vojta, M. & Wölfle, P. Fermi-liquid instabilities at magnetic quantum phase transitions. *Reviews of Modern Physics* **79**, 1015-1075 (2007).
7 Paschen, S. & Si, Q. Quantum phases driven by strong correlations. *Nature Reviews Physics* **3**, 9-26 (2020).
8 Varma, C. M. Colloquium: Linear in temperature resistivity and associated mysteries including high temperature superconductivity. *Reviews of Modern Physics* **92** (2020).
9 Ghiotto, A. *et al.* Quantum criticality in twisted transition metal dichalcogenides. *Nature* **597**, 345-349 (2021).
10 Nakatsuji, S. *et al.* Superconductivity and quantum criticality in the heavy-fermion system β-YbAlB4. *Nature Physics* **4**, 603-607 (2008).
11 Tomita, T., Kuga, K., Uwatoko, Y., Coleman, P. & Nakatsuji, S. Strange metal without magnetic criticality. *Science* **349**, 506-509 (2015).
12 Pfleiderer, C., Julian, S. R. & Lonzarich, G. G. Non-Fermi-liquid nature of the normal state of itinerant-electron ferromagnets. *Nature* **414**, 427-430 (2001).
13 Ritz, R. *et al.* Formation of a topological non-Fermi liquid in MnSi. *Nature* **497**, 231-234 (2013).
14 Smith, R. P. *et al.* Marginal breakdown of the Fermi-liquid state on the border of metallic ferromagnetism. *Nature* **455**, 1220-1223 (2008).
15 Jaoui, A. *et al.* Quantum critical behaviour in magic-angle twisted bilayer graphene. *Nature Physics* **18**, 633-638 (2022).
16 Burg, G. W. *et al.* Correlated Insulating States in Twisted Double Bilayer Graphene. *Physical Review Letters* **123** (2019).
17 Cao, Y. *et al.* Strange Metal in Magic-Angle Graphene with near Planckian Dissipation. *Physical Review Letters* **124** (2020).
18 Cao, Y. *et al.* Correlated insulator behaviour at half-filling in magic-angle graphene superlattices. *Nature* **556**, 80-84 (2018).
19 Chen, S. *et al.* Electrically tunable correlated and topological states in twisted monolayer–bilayer graphene. *Nature Physics* **17**, 374-380 (2020).
20 Cao, Y. *et al.* Unconventional superconductivity in magic-angle graphene superlattices. *Nature* **556**, 43-50 (2018).
21 Cao, Y. *et al.* Tunable correlated states and spin-polarized phases in twisted bilayer–bilayer graphene. *Nature* **583**, 215-220 (2020).
22 Liu, X. *et al.* Tunable spin-polarized correlated states in twisted double bilayer graphene. *Nature* **583**, 221-225 (2020).
23 Shen, C. *et al.* Correlated states in twisted double bilayer graphene. *Nature Physics* **16**, 520-525 (2020).
24 Lu, X. *et al.* Superconductors, orbital magnets and correlated states in magic-angle bilayer

graphene. *Nature* **574**, 653-657 (2019).
25 Park, J. M., Cao, Y., Watanabe, K., Taniguchi, T. & Jarillo-Herrero, P. Tunable strongly coupled superconductivity in magic-angle twisted trilayer graphene. *Nature* **590**, 249-255 (2021).
26 Xu, S. *et al.* Tunable van Hove singularities and correlated states in twisted monolayer–bilayer graphene. *Nature Physics* **17**, 619-626 (2021).
27 Yankowitz, M. *et al.* Tuning superconductivity in twisted bilayer graphene. *Science* **363**, 1059-1064 (2019).
28 Hao, Z. *et al.* Electric field–tunable superconductivity in alternating-twist magic-angle trilayer graphene. *Science* **371**, 1133-1138 (2021).
29 Cao, Y. *et al.* Nematicity and competing orders in superconducting magic-angle graphene. *Science* **372**, 264-271 (2021).
30 Chou, Y.-Z. & Das Sarma, S. Kondo Lattice Model in Magic-Angle Twisted Bilayer Graphene. *Physical Review Letters* **131**, 026501 (2023).
31 Song, Z.-D. & Bernevig, B. A. Magic-Angle Twisted Bilayer Graphene as a Topological Heavy Fermion Problem. *Physical Review Letters* **129**, 047601 (2022).
32 Checkelsky, J. G., Bernevig, B. A., Coleman, P., Si, Q. & Paschen, S. Flat bands, strange metals and the Kondo effect. *Nature Reviews Materials* (2024).
33 Ramires, A. Twisted-graphene model draws inspiration from heavy elements. *Nature* **608**, 474-475 (2022).
34 Ramires, A. & Lado, J. L. Emulating Heavy Fermions in Twisted Trilayer Graphene. *Physical Review Letters* **127** (2021).
35 Zhao, W. *et al.* Gate-tunable heavy fermions in a moiré Kondo lattice. *Nature* **616**, 61-65 (2023).
36 Zhao, W. *et al.* Emergence of ferromagnetism at the onset of moiré Kondo breakdown. *Nature Physics* **20**, 1772-1777 (2024).
37 Holleis, L. *et al.* Fluctuating magnetism and Pomeranchuk effect in multilayer graphene. *Nature* **640**, 355-360 (2025).
38 He, M. *et al.* Symmetry breaking in twisted double bilayer graphene. *Nature Physics* **17**, 26-30 (2021).
39 Li, Q. *et al.* Tunable quantum criticalities in an isospin extended Hubbard model simulator. *Nature* **609**, 479-484 (2022).
40 Han, T. *et al.* Correlated insulator and Chern insulators in pentalayer rhombohedral-stacked graphene. *Nature Nanotechnology* **19**, 181-187 (2023).
41 Abrahams, E., Kravchenko, S. V. & Sarachik, M. P. Metallic behavior and related phenomena in two dimensions. *Reviews of Modern Physics* **73**, 251-266 (2001).
42 Efros, A. L. & Shklovskii, B. I. Coulomb gap and low temperature conductivity of disordered systems. *Journal of Physics C: Solid State Physics* **8**, L49 (1975).
43 Chu, Y. *et al.* Temperature-linear resistivity in twisted double bilayer graphene. *Physical Review B* **106**, 035107 (2022).
44 Polshyn, H. *et al.* Large linear-in-temperature resistivity in twisted bilayer graphene. *Nature Physics* **15**, 1011-1016 (2019).
45 Bulla, R., Costi, T. A. & Pruschke, T. Numerical renormalization group method for quantum impurity systems. *Reviews of Modern Physics* **80**, 395-450 (2008).

46 Hewson, A. C. *The Kondo problem to heavy fermions*. (Cambridge university press, 1997).

47 Khadka, D. *et al.* Kondo physics in antiferromagnetic Weyl semimetal $Mn_{3+x}Sn_{1-x}$ films. *Science Advances* **6**, eabc1977.

48 Jang, S. *et al.* Evolution of the Kondo lattice electronic structure above the transport coherence temperature. *Proceedings of the National Academy of Sciences* **117**, 23467-23476 (2020).

49 Sumiyama, A. *et al.* Coherent Kondo State in a Dense Kondo Substance: $Ce_xLa_{1-x}Cu_6$. *Journal of the Physical Society of Japan* **55**, 1294-1304 (1986).

50 Perkins, N. B., Iglesias, J. R., Núñez-Regueiro, M. D. & Coqblin, B. Coexistence of ferromagnetism and Kondo effect in the underscreened Kondo lattice. *Europhysics Letters (EPL)* **79**, 57006 (2007).

51 Lavagna, M., Lacroix, C. & Cyrot, M. Resistivity of the Kondo lattice. *Journal of Physics F: Metal Physics* **12**, 745 (1982).

52 Chowdhury, D., Georges, A., Parcollet, O. & Sachdev, S. Sachdev-Ye-Kitaev models and beyond: Window into non-Fermi liquids. *Reviews of Modern Physics* **94**, 035004 (2022).

53 Lee, S.-S. Recent Developments in Non-Fermi Liquid Theory. *Annual Review of Condensed Matter Physics* **9**, 227-244 (2018).

54 Watanabe, H. & Vishwanath, A. Criterion for stability of Goldstone modes and Fermi liquid behavior in a metal with broken symmetry. *Proceedings of the National Academy of Sciences* **111**, 16314-16318 (2014).

55 Hofstetter, W. & Schoeller, H. Quantum Phase Transition in a Multilevel Dot. *Physical Review Letters* **88**, 016803 (2001).

56 Žitko, R. & Bonča, J. Enhanced conductance through side-coupled double quantum dots. *Physical Review B* **73**, 035332 (2006).